\documentclass[11pt,a4paper]{article}
\usepackage{jcappub}

\usepackage[utf8x]{inputenc}
\usepackage[UKenglish]{babel}
\usepackage{amsmath}
\usepackage{amssymb}
\usepackage{amsfonts}
\usepackage{graphics}
\usepackage[normal]{subfigure}
\usepackage{color}
\usepackage{tensor}
\usepackage{amsmath}

\def\be{\begin{equation}}
\def\ee{\end{equation}}

\newcommand{\phienics}{$\varphi$\texttt{enics}}
\newcommand{\fenics}{\texttt{FEniCS}\ }

\newcommand{\rSrc}{\ensuremath{r_{\rm s}}}
\newcommand{\mSrc}{\ensuremath{M_{\rm S}}}
\newcommand{\stepWid}{\ensuremath{w}}

\title{Massive Galileons and Vainshtein Screening}
\author[a]{Clare Burrage,}
\author[a]{Ben Coltman,}
\author[a]{Antonio Padilla,}
\author[a]{Daniela Saadeh,}
\author[a]{and Toby Wilson}
\affiliation[a]{School of Physics and Astronomy, University of Nottingham,\\
 Nottingham NG7 2RD, United Kingdom}
\emailAdd{clare.burrage@nottingham.ac.uk}
\emailAdd{ben.coltman@nottingham.ac.uk}
\emailAdd{antonio.padilla@nottingham.ac.uk}
\emailAdd{daniela.saadeh@nottingham.ac.uk}
\abstract{The Vainshtein screening mechanism relies on nonlinear interaction terms becoming dominant close to a compact source. However, theories displaying this mechanism are generally understood to be low-energy theories: it is unclear that operators emerging from UV completion do not interfere with terms inducing Vainshtein screening. In this work, we find a set of interacting massive Galileon theories that exhibit Vainshtein screening; examining potential UV completions of these theories, we determine that the screening does not survive the extension. We find that neglecting operators when integrating out a heavy field is non-trivial, and either care must be taken to ensure that omitted terms are small for the whole domain, or one is forced to work solely with the UV theory.  We also comment on massive deformations of the familiar Wess-Zumino Galileons. }
\begin{document}
\maketitle

\section{Introduction} \label{intro}

There is now a wealth of observational evidence in support of an accelerating  universe \cite{SN1,SN2, CMB}.  This acceleration is usually attributed to a cosmological fluid known as dark energy whose microscopic origins are unknown \cite{ed}  or else some modification of Einstein's General Theory of Relativity at large distances \cite{review1,review2, Ishak_review}. In some of the most interesting phenomenological scenarios, this acceleration can be identified with the dynamics of an ultra-light scalar field which couples to ordinary matter with gravitational strength. If the scalar continued to operate in this way at shorter distances - within the scale of the solar system - it  would mediate a fifth fundamental force that so far has not been detected \cite{Adelberger:2003zx,Adelberger:2006dh}. Viable models must therefore be able to screen, i.e. suppress, the extra force in environments where it is known to be small. Only a handful of screening mechanisms are known (see eg \cite{Dehnen:1992rr, Gessner:1992flm, Damour:1994zq, Pietroni:2005pv, Olive:2007aj,cham1,cham2, sym1,sym2, Vainshtein}), one of which is Vainshtein screening (for a review see \cite{vainrev}). Here, a derivative interaction term dominates close to a matter source, causing a breakdown of the linear theory and suppressing the gradient of the scalar field, thus screening the fifth force within a typically large Vainshtein radius. Vainshtein screening is seen in nonlinear massive gravity \cite{NLMGexample} and Galileon-type models \cite{galexample}. Theories displaying Vainshtein screening necessarily run into strong coupling at macroscopic scales in order for the derivative interactions to kick in at sufficiently large distances from the source \cite{power}.  For this reason, these theories can only be properly understood as effective theories with a limited range of validity. Since the breakdown occurs on macroscopic scales it is important to ask what happens beyond that scale and what impact it has on Vainshtein screening. This question has been studied before \cite{NK,ippo}, where it was argued that a generic ultra-violet (UV) completion of a theory with derivative interactions could introduce further interaction terms that have the potential to destroy the Vainshtein mechanism. However, the difficulty in addressing this question directly has been the absence of a known UV  completion of a theory that exhibits Vainshtein screening  (indeed, in the case of Galileons \cite{galexample}, it has been argued that a standard Wilsonian UV completion does not exist \cite{posbounds1}).  

\begin{figure}[h]
\centering
\includegraphics[width=\columnwidth]{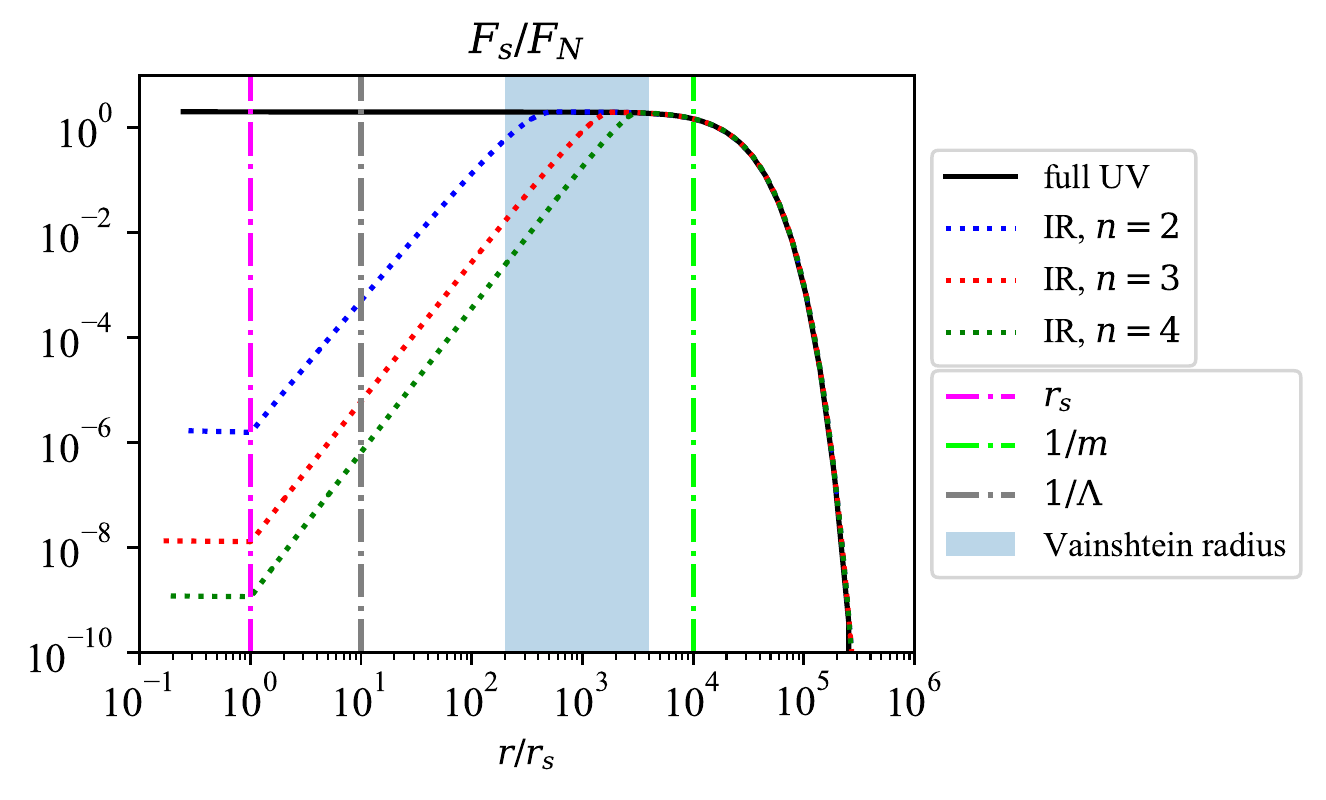}
\caption{Ratio of the scalar fifth force to Newtonian force around a compact object, for the IMG  theories described in the text (Eq.~\ref{EOM1}) (dotted lines) and a possible UV completion for the $n=3$ case (Eq.~\ref{EOM2}) (solid black line). The shaded region indicates the distance scales where the behaviour of the force changes for the different theories. When pushed towards strong coupling, the IMGs show marked suppression of the fifth force around the compact object. However, because this occurs at strong coupling, one really ought to work with a description that extends the theory further into the UV. For the UV complete example shown here (solid black line), we see that there  is no longer any suppression  of the fifth force. }
\label{Fig:force_plot}
\end{figure}

In this paper, we examine the potential of Vainshtein screening to survive UV completion directly.  This has been made possible with the advent of an interacting massive Galileon (IMG) and its UV completion presented in \cite{MassGals}.  Motivated by that set-up we examine a generalised  set of IMG  theories  together with their  possible UV completions.  We show that Vainshtein screening does occur for each type of interaction provided the Galileon is massive. Armed with an extended description at high energies, we are able to see if screening survives the inclusion of  UV corrections. The answer is a resounding no.  Through these explicit examples, it becomes clear that a low energy approximation to any UV theory is not automatically trustworthy when pushed into a non-perturbative regime. Such conclusions should not come as a surprise given our understanding of effective field theories in particle physics. Nevertheless, the conclusion is significant in the context of Vainshtein screening, where at least some higher order operators are required to become large by construction. These results cast  further doubt on the theoretical viability of Vainshtein screening, even before observational constraints are attempted.

Our approach combines analytic estimates with a careful numerical analysis.  A flavour of the numerical results are presented in Fig.\ref{Fig:force_plot} where we plot the ratio of the fifth force to the standard Newtonian force in the vicinity of a spherically symmetric compact source.  The dotted lines reveal what happens for a family of massive Galileon theories with particular derivative interactions. In each case, the fifth force is suppressed close to the source  as the derivative interaction begins to dominate.  The black solid line is the prediction for a UV completion of one of these scenarios. It is easy to see that suppression of the fifth force no longer occurs: the Vainshtein mechanism is completely destroyed by the UV corrections to the theory.

The rest of this paper is organised as follows: in the next section, we will identify a family of higher order Galileon invariant operators that have the potential to Vainshtein screen, but {\it only} when the Galileon is massive.  We give simple analytic arguments to indicate that screening will take place which are then reinforced by our numerical analysis.  In section \ref{sec:raise} we raise the cut-off of our effective description  by integrating in a heavy field.  This is done for each family of interactions considered in section \ref{IR}.   We present a generic analytic argument for why  we expect screening to be spoiled in these UV extended  theories. For the special case already identified in \cite{MassGals}, the theory in question is UV complete in the limit $M_\text{Pl} \to \infty$.  In section \ref{sec:num}, we perform a numerical analysis on this UV complete theory and see that screening is destroyed.  This allows us to scrutinise the integrating 
 process in detail, and re-examine operators one would usually neglect due to heavy mass suppression.  It turns out that a tower of higher order operators can no longer be neglected within a certain macroscopic distance from the source.  This is entirely consistent with the generic arguments presented in \cite{NK,ippo} and reinforces the idea that Vainshtein screening cannot be taken seriously without a much better understanding of the UV effects in any particular model. Our numerical methods and results are presented in section \ref{sec:num}, with additional details found in a companion paper \cite{method_paper}.  We conclude  in section \ref{sec:conc}.   In the appendix, we consider adding a mass deformation to  so-called Wess-Zumino Galileon theories, and ask whether the screening properties remain intact. 

We work in units with $c=\hbar=1$, and use the reduced Planck mass $M_\text{Pl}=1/\sqrt{8\pi G}$.

\section{Interacting Massive Galileons and Vainshtein screening} \label{IR}
 Galileon theories \cite{galexample} have been seen to emerge in a variety of interesting cosmological scenarios, from DGP gravity \cite{DGP} to non-linearly realised massive gravity \cite{DRGT}.  Although they contain higher order derivative interactions, the field equations remain at second order, thereby avoiding the Ostrogradski instability \cite{ostro}.  The defining characteristic of a Galileon theory is one that is invariant  under a Galileon transformation $\pi\rightarrow \pi + b_\mu x^\mu + c$ in flat space, where $b_\mu$ and $c$ are constant and we sum repeated indices following the Einstein conventions.  For the theories with second order field equations defined  in \cite{galexample}, the interaction operators shift by a total derivative under the Galileon transformation and for this reason they are sometimes referred to as Wess-Zumino interactions \cite{Goon}. Of course, we can also include interactions that are {\it manifestly} invariant under the Galileon transformation, such as $(\partial  \partial  \pi )^n$, where there are two (or more) derivatives acting on each insertion of the scalar. These are expected to arise anyway as effective field theory (EFT) corrections to the leading order interactions.

The Wess-Zumino interactions are known to facilitate Vainshtein screening \cite{galexample}. Although, as emphasised in the introduction, this goes hand in hand with strong coupling  and concerns about the validity of our effective description when screening is active \cite{NK, ippo}.   One way to avoid this concern would be to find a UV theory which could reproduce the Vainshtein mechanism at low energies, but generalises to higher energies. Unfortunately, positivity constraints suggest that a standard Wilsonian UV completion of the theory cannot exist \cite{posbounds1, posbounds2}.  To avoid these bounds, it is necessary to deform the Galileon theory in the infra-red (IR), in order to change the form of the low energy scattering amplitudes. One of the simplest deformations is the inclusion of a mass term: although such a term breaks the Galileon symmetry, the action continues to respect the Galileon non-renormalisation theorem at low scales \cite{Burrage:2011cr}. Additionally, loop corrections generated by the addition of a mass term do not violate the Galileon symmetry at any order. We then look to be in good shape to have a well-behaved theory that has some hope of being UV completed.

An example of a UV-complete massive Galileon theory was given in Ref.~\cite{MassGals}, where, through the introduction of a single heavy field $H$, Galileon invariant interactions for the light Galileon field $\pi$ can be obtained, with the exception of the mass term. Integrating out the heavy field to leading order\footnote{We do this by substituting the equation of motion for $H$ back into the action. We explicitly computed the first-order loop corrections to these results and they do not alter the form of the would-be screening operator.} yields a single field Galileon theory in $\pi$  at low energies, respecting  the same symmetry. Generalising the self interaction term for the heavy field to any integer power $n+1$, the IR theory then contains terms of the form $(\Box \pi)^n$. Further details on how this is done can be found in section \ref{sec:raise}. The non-linear nature of the derivative interactions opens up the possibility that screening will occur.

With this in mind, we consider the following action assumed to be valid at low energies\footnote{Note that we employ a different definition of the parameters $\epsilon$ and $\Lambda$ compared to those of the companion paper \cite{method_paper}.}, 
\begin{equation} \label{action1}
S[\pi]=\int d^4x \left( -\frac{1}{2} \left(\partial \pi \right)^2-\frac{1}{2}m^2 \pi ^2 +\frac{\epsilon}{n+1} \frac{(\Box \pi)^{n+1}}{\Lambda^{3n-1}} + \frac{\pi T}{M_\text{Pl}} \right) \end{equation}
where $\pi$ is a scalar field with mass $m$, the integer $n\in \{2,3,4\ldots\}$ and $\epsilon=\pm 1$. One can see that this action is invariant under the Galileon transformation of $\pi\rightarrow \pi + b_\mu x^\mu + c$, with the exception of the mass term, verifying that it is indeed a theory of a massive Galileon\footnote{For clarity and brevity, we define a massive Galileon in flat space as a theory for which $\delta \mathcal{L} \propto m^2 $ under the Galileon transformation, up to total derivatives.}. As usual, the Galileon is coupled to external sources with gravitational strength through the trace of the energy-momentum tensor $T$.  The theory becomes strongly coupled at some scale $\Lambda \ll M_\text{Pl}$, reflecting its status as an effective theory only valid at large distances. We now ask the following: does this theory exhibit Vainshtein screening close to the source and, if so,  how close to the source can we go and still trust its predictions? The latter requires knowledge of the UV completion to be discussed in the next section.  

We proceed by varying the action to obtain the equation of motion,
\begin{equation} \label{EOM1}
\Box \pi -m^2 \pi + \epsilon \mathcal{O}_n = \frac{M_\text{S}}{M_\text{Pl}} \delta (\underline{x})
\end{equation}
where $\mathcal{O}_n \equiv \frac{\Box(\Box \pi)^n}{\Lambda^{3n-1}}$, and we have chosen a pressureless delta function source of mass $M_\text{S}$ with support at $\underline{x}=\underline{0}$.  We shall now look for static, spherically symmetric configurations.

Firstly, considering Eq. (\ref{EOM1}) far from the source, we are in the so-called linear regime, and the solution has the form $\pi_\text{lin} \sim \frac{M_\text{S}}{M_\text{Pl}} \frac{e^{-mr}}{r}$. In order to determine at what radius we might expect a breakdown of the linearised theory, and therefore identify a candidate Vainshtein radius, we evaluate $\mathcal{O}_n$ on the linearised solution, and compare it to the other terms in the equation of motion. We find that $\mathcal{O}_n|_{\pi_\text{lin}} \sim \Lambda^{1-3n} m^{2n} \Box \pi_\text{lin}^n$. Assuming that we are well inside the Compton wavelength of $\pi$, we can take the approximation $r \ll m^{-1}$, which then simplifies the expression to $\mathcal{O}_n|_{\pi_\text{lin}} \sim \Lambda^{1-3n} m^{2n}  \left(\frac{M_\text{S}}{M_\text{Pl}}\right)^n r^{-(n+2)}$.

Comparing with the mass term, the ratio $\mathcal{O}_n|_{\pi_\text{lin}}/m^2 \pi_\text{lin}$ is given by $\left(\frac{r_v^{(n)}}{r}\right)^{n+1}$, where $r_v^{(n)} \sim (\sigma_\text{S} \kappa^2)^\frac{n-1}{n+1} \Lambda^{-1}$, with $\sigma_\text{S} \equiv \frac{M_\text{S}}{M_\text{Pl}}$ and $\kappa \equiv \frac{m}{\Lambda}$. We see that so long as $\sigma_\text{S} \kappa^2 \gg 1$, then the linearised theory breaks down at some macroscopic scale $r_v^{(n)} \gg \Lambda^{-1}$. It is worth recognising that, without a mass term, $\kappa=0$ and there is no screening.

Although we have identified a potential breakdown of the linear theory, we still have not confirmed the existence of screening; we must examine the non-linear regime and determine whether the solution supports a screening mechanism. To this end, we neglect the kinetic and mass terms in (\ref{EOM1}), and integrate the equation to obtain $\frac{(\Box \pi)^n}{\Lambda^{3n-1}} \sim \frac{M_\text{S}}{M_\text{Pl}} \frac{1}{r} + c$ where $c$ is a constant. If the constant is negligible, we integrate to obtain a solution of the form $\pi \sim (\sigma_\text{S} \Lambda^{3n-1})^{\frac{1}{n}} r^{2-\frac{1}{n}}$. However, if the constant instead dominates, the solution is of the form $\pi \sim (c  \Lambda^{3n-1})^{\frac{1}{n}} r^2 + d$. We see that in both cases the scalar force is suppressed at small radii, consistent with screening. 

To complete our analysis, we need to show that the two asymptotic solutions, at large and small radii, can be consistently matched onto one another. We have not been able to show this analytically, but our numerical solutions indicate that the two solutions can indeed be matched (see Fig.\ref{Fig:force_plot}). This suggests that the family of interacting massive Galileon theories given by equation \eqref{action1} will exhibit Vainshtein screening around a heavy source. However, given the importance of the derivative interaction in suppressing the force close to the source, it remains to ask whether or not we really trust this prediction.  UV corrections are expected in order to preserve perturbative unitarity and raise the cut-off of the effective theory. What effect do these corrections have on the predictions of the theory close to the source?

\section{Raising the cut-off eliminates screening} \label{sec:raise}

Consider the action,
\begin{align} 
   S[\pi ,H]=\int d^4x& \left( -\frac{1}{2} \left(\partial \pi \right)^2 -\frac{1}{2}\left(\partial H\right)^2-\frac{1}{2}m^2 \pi ^2 \right.\nonumber
   \\
   &\;\;\;\left. -\frac{1}{2} M^2 H^2 -\alpha  H \square \pi -\frac{ \lambda H^{n+1}}{(n+1)!\mu^{n-3}} + \frac{\pi T}{M_\text{Pl}} \right)\label{action2}
\end{align}
generalised from \cite{MassGals}, where $\pi$ is the Galileon field, with light mass $m$, $H$ is some heavy field of mass $M\gg m$, and $T$ is the trace of the energy-momentum tensor of the source, coupling only to the Galileon field.  $\lambda$ and $\alpha$ are dimensionless coupling coefficients of order one, although we must impose $\lambda \geq 0$ and $|\alpha|<1$  to avoid instabilities.  $\mu$ is a new high energy scale representing the new cut-off of the theory when $n\geq4$.  For $n \in \{2,3\}$, the theory is well-defined all the way up to the Planck scale. One can see that in each case the action will transform in the correct way in order to be considered a massive Galileon theory.   Variation yields the following field equations,
\begin{align}
\left\lbrace
\begin{array}{l}
\Box \pi -m^2 \pi - \alpha \Box H = -\frac{T}{M_\text{Pl}} \\
\Box H -M^2 H - \alpha \Box \pi -\frac{ \lambda  H^n}{n!\mu^{n-3}} = 0\;.
\end{array} \right.
 \label{EOM2}
\end{align}
We assume as boundary conditions that the fields are everywhere regular and asymptoting to the vacuum expectation value.

Examining the region far outside the Compton wavelength of $H$, we can make the assumption $\Box \ll M^2$, which gives
\begin{equation}
H \sim -\frac{\alpha}{M^2} \Box \pi - \frac{\lambda \alpha^{n} (-1)^n}{n! M^{2(n+1)}} \frac{(\Box \pi)^n}{\mu^{n-3}} + O(\lambda^2)\;.
\end{equation}
It should be noted that we have discarded terms of the form $\left(\frac{\Box}{M^2}\right)^j \pi$ in order to write down this expression. While these are legitimate terms under all of our perturbation expansions, they are subdominant in both the linear and non-linear regimes, and only become important at the Compton wavelength of $H$, at which point one would have to work with the full UV theory anyway.

Proceeding, we can write a low energy action as,
\begin{equation}
\widetilde{S} = \int d^4x  \left( -\frac{1}{2} \left(\partial \pi \right)^2  -\frac{1}{2}m^2 \pi ^2 + \frac{(-1)^n \lambda \alpha^{n+1} (\Box \pi)^{n+1}}{(n+1)! M^{2(n+1)}\mu^{n-3}} + \frac{\pi T}{M_\text{Pl}}\right)
\end{equation}
where we have discarded terms of order $\lambda^2$ or higher. We then identify this with our IR theory, and see that we must have $M^{2(n+1)}\mu^{n-3} \sim \Lambda^{3n-1}$ and, for $n$ odd, $\epsilon=-1$. If we want the UV theory to be able to describe physics at higher energies reliably, we require it to have a larger strong coupling scale than the corresponding IR theory. For $n\in \{2,3\}$, the UV theory is renormalisable in the absence of external sources, but for $n\geq 4$ we must restrict ourselves to $\Lambda<\mu$. Writing $\mu = N \Lambda$ for $N>1$, we see that $M=N^\frac{3-n}{2(n+1)} \Lambda$, i.e. the heavy field must be lighter than the strong coupling scale, in keeping with our intuition from Wilsonian UV completions.

We now give analytic arguments to explain why we expect screening to be absent in this extended theory, focussing on the UV complete case with $n=3$. We start by rewriting the equations of motion as follows:
\begin{align}
\left\lbrace
\begin{array}{l}
\Box (\pi-\alpha H) - m^2 \pi = \frac{\rho}{M_\text{Pl}} \\
\Box (H-\alpha \pi) - V'(H) = 0
\end{array} \right.
\label{Eq:analytics_EOM}
\end{align}
where $V'(H) = M^2 H + \frac{\lambda}{3!}H^3$, and for simplicity the source $\rho$ is taken to be a top-hat function of radius $r_\text{s}$, i.e. $\rho(r)=\bar{\rho}\,\Theta(r_\text{s} - r)$, so that we may explore the field profiles both inside and outside the source. The main focus here will be on the solution for the Galileon field, $\pi$, since this is the one probed directly by matter.

We start by assuming that $\beta \equiv \Box H/V'(H)$ varies slowly. This is consistent with the numerical simulations everywhere away from the source-vacuum transition. In principle the constant value of $\beta$ could differ from inside to outside the source.  The second equation in Eq. \eqref{Eq:analytics_EOM} now yields
 $\Box H=\frac{\alpha}{1-\beta^{-1}} \Box \pi$ and substituting this into the first equation  gives
\begin{equation}
(Z\Box - m^2)\pi = \frac{\rho}{M_\text{Pl}}
\end{equation}
where $Z\equiv1-\frac{\alpha^2}{1-\beta^{-1}}$ is assumed to be positive.  It is convenient to define effective mass scales $\bar m_\text{in}=m/\sqrt{Z_\text{in}}$ and $\bar m_\text{out}=m/\sqrt{Z_\text{out}}$ so that  this equation has the regular solution:

\begin{align}
\pi_\text{in}(r) = & - \frac{\bar{\rho} }{M_\text{Pl} m^2} \left[ 1-  \frac{(1+x_\text{out}) \sinh(\bar m_\text{in} r ) }{x_\text{out} \sinh x_\text{in}+x_\text{in} \cosh x_\text{in}}\frac{r_s}{r} \right] \\
\begin{split}
 \pi_\text{out}(r) =  & -\frac{ \bar{\rho} }{M_\text{Pl} m^2}   \left[   \frac{ e^{x_\text{out}} (x_\text{in} \cosh x_\text{in}- \sinh x_\text{in}) }{x_\text{out} \sinh x_\text{in}+x_\text{in} \cosh x_\text{in}} \right] \frac{r_s}{r} e^{-\bar m_\text{out} r }
\end{split}
\end{align}
where  we define $x_\text{in}\equiv x/\sqrt{Z_\text{in}}$ and  $x_\text{out}\equiv x/\sqrt{Z_\text{out}}$ for  $x \equiv m r_s$.  Note that the solutions match at the source-vacuum transition, along with their first derivatives.  We will also assume that the source lies deep within the Compton wavelength of the Galileon, so in other words, $x \ll 1$. To examine screening, we compare the exterior solution $\pi_\text{out}$ with a typical Newtonian potential, $V_N =- \frac{\bar{\rho} r_\text{s}^3 }{6 {M_\text{Pl}}^2 r} $. The ratio
\begin{equation}
\frac{\pi_\text{out}/M_\text{Pl}}{V_N}=\frac{6}{x^2}  \left[   \frac{ e^{x_\text{out}} (x_\text{in} \cosh x_\text{in}- \sinh x_\text{in}) }{x_\text{out} \sinh x_\text{in}+x_\text{in} \cosh x_\text{in}} \right] e^{-\bar m_\text{out} r }
\end{equation}
is suppressed in two cases. The first corresponds to Yukawa suppression in the exterior, with $Z_\text{out} \ll 1$. Alternatively, if $Z_\text{out} \gtrsim  1$, suppression can also occur if the scalar decouples in the interior, with $Z_\text{in} \gg 1$. 
We shall now demonstrate that these scenarios are incompatible with the required profile for $H$ and so screening is not possible, at least up to the caveat of our approximations. 

Recall that $\Box H=\frac{\alpha}{1-\beta^{-1}} \Box \pi=\frac{(1-Z)}{\alpha}\Box \pi$ and so $H=\frac{(1-Z)}{\alpha} \pi +\hat H$ where $\Box \hat H=0$.  Assuming regularity and continuity of $H$ and its first derivative at the transition, we obtain
\begin{align}
H_\text{in} = & \frac{(1-Z_\text{in})}{\alpha} \pi_\text{in}+ \frac{(Z_\text{out}-Z_\text{in})}{\alpha} x_\text{out} \pi_\text{s} \\
H_\text{out} =&\frac{(1-Z_\text{out})}{\alpha} \pi_\text{out}+\frac{(Z_\text{out}-Z_\text{in})}{\alpha} (x_\text{out}+1)\pi_\text{s} \frac{r_\text{s}}{r} 
\end{align}
where $\pi_\text{s} \equiv -\frac{\bar{\rho}}{M_\text{Pl} m^2} $ $\left[  \frac{x_\text{in} \cosh x_\text{in}- \sinh x_\text{in} }{x_\text{out} \sinh x_\text{in}+x_\text{in} \cosh x_\text{in}} \right]$ is the value of the Galileon at the transition. 

For the case of Yukawa suppression for the exterior Galileon, we have $Z_\text{out} \ll 1$. The Yukawa suppression allows us to neglect $\pi_\text{out}$ in $H_\text{out}$. This means that $H_\text{out}$ scales like a massless field in most of the exterior, and given our definition $\beta \equiv \Box H/V'(H)$, we infer $\beta_\text{out} \ll 1$. The problem now is that this gives $Z_\text{out} \approx 1$ in contradiction with the condition for Yukawa suppression. 

For the case of suppression through decoupling of the interior Galileon, we have $Z_\text{in} \gg 1$. It follows that $\pi_\text{in} \approx -\frac{\bar{\rho}}{6 M_\text{Pl} Z_\text{in}} \left( \frac{3+x_{\rm out}}{1+x_{\rm out}} {r_\text{s}}^2 - r^2 \right)$ and so $\Box H_\text{in} \approx -\frac{\bar{\rho}}{M_\text{Pl} 
\alpha}$.  However, for $Z_\text{in} \gg 1$ we require $\beta_\text{in} \approx 1$, and so we now expect $V'(H_\text{in}) \approx  -\frac{\bar{\rho}}{M_\text{Pl} \alpha}$. This suggests $H_\text{in} \approx $ constant, in obvious contradiction with $\Box H_\text{in} \approx -\frac{\bar{\rho}}{M_\text{Pl} 
\alpha}$, except in the trivial limit where $\bar{\rho} \to 0$.

In summary then, our heuristic analysis seems to suggest that screening of the Galileon will not be possible when the backreaction of the heavy field is taken into account. Of course, the assumption of constant $\Box H/V'(H)$ was a little crude and the numerics show that this does not hold particularly well near the vacuum-source transition, casting some doubts on our right to apply continuity conditions at this point.  For these reasons we do not present our analytics as the main evidence that ultra-violet effects will spoil the Vainshtein effects. We leave that to the numerics.

\section{Numerical methods and results} \label{sec:num}

Determining the screening property of the UV theory analytically is challenging if one is to avoid some crude assumptions. Likewise, for the IR theory, there is no a-priori guarantee that it is possible to match between the high- and low-density regimes consistently. We therefore address the problem numerically, to obtain the solution to the full equations of motion Eq.~\eqref{EOM1}, for $n=2,3,4$ and Eq.~\eqref{EOM2} for $n=3$ across all regimes. For this task, we have developed the numerical code \phienics\footnote{\protect\url{https://github.com/scaramouche-00/phi-enics}} \cite{method_paper}, building on the \fenics library\cite{OldFenicsbook,Fenicscitations,NewFenicsbook}. \phienics{} applies the finite element method to the solution of boundary-value problems relevant for screening, and is able compute the fields' profiles, associated fifth force and high-order operators accurately across the full simulation box, without restricting to the high- and low-density regimes to which analytic understanding is generally confined. The finite element method is well suited for the computation of the high-order operators $\Box(\Box\pi)^n$ under study, for which traditionally employed finite-differencing techniques are not sufficient.

For both theories, we compute the field profiles in the presence of a static spherically symmetric compact source of mass $\mSrc=10^{10} M_\text{Pl}$ and radius $\rSrc=10^{47} {M_\text{Pl}}^{-1}$, following a smoothed top-hat profile:
\begin{equation}
\rho(r) =  \frac{\mSrc}{ 4 \pi(-2\stepWid^3)\textrm{Li}_3(-e^{\bar{r}/\stepWid}) } \frac{1}{\exp{\frac{r - \bar{r}}{ \stepWid }} + 1 } 
\label{Eq:Source}
\end{equation}
where $\stepWid=0.02 \rSrc$, $\textrm{Li}_3(x)$ is the polylogarithm function of order 3 and $\bar{r}$ is chosen so that 95\% of the source mass is included within $\rSrc$. In the limit $w/\rSrc \rightarrow 0$, this density profile becomes the step function $\rho(r)=0.95\frac{3\mSrc}{4\pi\rSrc^3}\Theta(\bar{t}\rSrc-r)$, where ${\bar{t}}^{-1}=\sqrt[3]{0.95}$ and $\Theta$ is the step function. We will find that the presence or absence of screening is not sensitive to the particular choice of smoothing we make to the step function density profile.  Whilst the specific choice of density profile may make minor changes to the behaviour of the field profile inside the source and close to the surface, it leaves the behaviour at larger radii unaffected \cite{method_paper}.

For the UV theory, we take the masses of the light and heavy fields to be $m=10^{-51} M_\text{Pl}$ and $M=10^{-48} M_\text{Pl}$, with coupling constants $\alpha=0.4$ and $\lambda=0.7$. For the IR theory, we take $\Lambda=2.07\times10^{-48}$ and $\epsilon=-1$. 
Note that this choice of parameters corresponds to different signs for $\alpha$ in the UV theory for $n=2,3,4$. 
For both theories, we impose that the fields be regular and asymptoting to the vacuum expectation value, which imposes the boundary conditions $\lbrace \phi(\infty)=H(\infty)=0; \nabla\phi(0)=\nabla H(0)=0 \rbrace$ and $\lbrace \pi(\infty)=0; \nabla\pi(0)=0; \nabla [\nabla^2\pi^n](\infty) = 0 \rbrace$. For the IR theory, we supplement these conditions with the requirement $\lbrace \nabla [\nabla^2\pi^n](0)=\mathrm{finite} \rbrace$, which is obtained from the numerical solution to the UV theory ($n=3$). The latter is applied for consistency with the requirement of UV completion.

We shall now give details of the settings used to solve the UV and IR theories. For both, we use interpolating polynomials of order $5$, and the following \phienics{} settings:
\begin{itemize}
    \item[--] UV theory, $n=3$: \texttt{ArcTanExpMesh} of $150$ points spanning a box $r \in [0, 10^{10}]\times \rSrc$, with parameters $k=8, a=5\times10^{-2}, b=3\times10^{-2}$. Field rescalings: $\mu_{\phi}=10^{13} M_\text{Pl}$, $\mu_H=10^{12} M_\text{Pl}$;
    
    \item[--] IR theory, $n=2$: \texttt{ArcTanExpMesh} of $400$ points, spanning a box $r \in [0, 10^{9}]\times \rSrc$, with parameters $k=25, a=5\times10^{-2}, b=3\times10^{-2}$ and declustering at $r_{\rm rm}=10^3 \rSrc$ with parameters $A_{\rm rm}=1, k_{\rm rm}=10$. Field rescaling: $\mu_{\pi}=10^{-15} M_\text{Pl}$.
    
    \item[--] IR theory, $n=3$:  \texttt{ArcTanExpMesh} of $700$ points spanning a box $r \in [0, 10^{9}]\times \rSrc$, with parameters $k=25, a=5\times10^{-2}, b=4\times10^{-2}$. Field rescaling: $\mu_{\pi}=10^{-15} M_\text{Pl}$.
    
    \item[--] IR theory, $n=4$: \texttt{ArcTanExpMesh} of $600$ points spanning a box $r \in [0, 10^{9}]\times \rSrc$, with parameters $k=25, a=5\times10^{-2}, b=3\times10^{-2}$. Field rescaling: $\mu_{\pi}=10^{-15} M_\text{Pl}$.

\end{itemize}

\noindent The mesh classes available in \phienics{} are discussed extensively in \cite{method_paper}: they apply a nonlinear transformation to a mesh that is initially equally spaced in order to obtain a discretisation that is finer along the source-vacuum transition and coarser everywhere else. All numerical settings reported here are similarly defined in \cite{method_paper} and in the \phienics{} documentation.

In Figure \ref{Fig:force_plot}, we show the ratio of the  scalar force to the Newtonian gravitational force $F_s/F_N$ for the UV theory ($n=3$) and the IR theory ($n=2,3,4$), around the compact object in Eq.~\eqref{Eq:Source}. When a scalar field couples to matter with a coupling strength $M_\text{Pl}$, the ratio is equal to $2$ if there is no screening. We can see that this is the case for the UV theory, where $F_s/F_N=2$ for $r \lesssim 1/m$, (for $r \gtrsim 1/m$ the massive field decays exponentially and the scalar force is correspondingly suppressed). The scenario is radically different for the IR theories, where strong Vainshtein screening is displayed around the source. Here, the scalar force is suppressed compared to the Newtonian force by a factor which can be as large as $10^9$, confirming our expectations of Sec.~\ref{sec:raise}.

\begin{figure}
\centering
\includegraphics{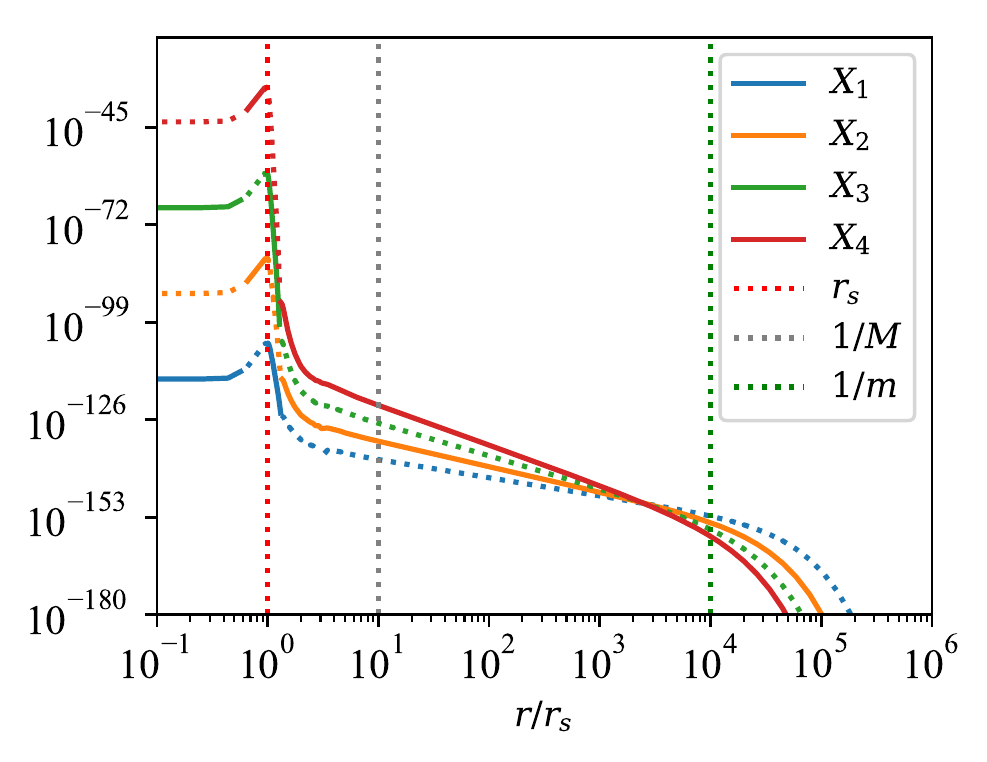}
\caption{The operators $X_j$ in Eq.~\eqref{Eq:Operators}, for the UV theory ($n=3$); solid (dotted) lines indicate positive (negative) values. The assumption $X_{j>1} \ll \mathcal{O}_3\equiv\frac{\Box(\Box\pi)^3}{\Lambda^8}$ (for $\Lambda^8= 6 M^8 / (\lambda\alpha^4)$) is clearly invalid.} 
\label{Fig:operators}
\end{figure}

To understand the absence of screening in the UV theory, and its apparent presence in the IR, we consider the neglected higher order terms for $n=3$.
Still under the assumption $\Box \ll M^2$, i.e. far from the Compton wavelength of $H$, we write down the leading order term for each power of $\lambda$, and find them to be of the form:
\begin{equation}
 X_j = (-1)^{j+1}  \binom{3j}{j} \frac{1}{2j+1} \alpha^{2j+2} \left(\frac{\lambda}{3!}\right)^j \Box(\Box \pi)^{2j+1} M^{-6j-2}
 \label{Eq:Operators}
\end{equation}
with $j\geq 1.$ We might, at first, expect
that the terms $j>1$ are negligible when compared to $X_1\equiv\mathcal{O}_3$ from Sec.~\ref{sec:raise}. However, when evaluated on the 
full UV solution,
 we find that actually these terms become important sooner than $\mathcal{O}_3$, and all at roughly the same radius. 
  We check this numerically, and compute the operators $X_j$ for $j=1,2,3,4$ in the UV theory ($n=3$): the result is shown in Figure~\ref{Fig:operators}. As expected, the hierarchy of the operators breaks down. We have verified that this numerical result is independent of the specific source profile or theory parameters used. Na\"{i}vely, we could consider the radius at which the higher-order operators become important as a new scale at which we might expect the linear theory to break down: however, this is not borne out by the numerical solution. It is therefore clear that the operators we initially neglected, along with $\mathcal{O}_3$, resum to produce an operator that is negligible and unable to provide screening at macroscopic distances.

\section{Conclusion} \label{sec:conc}
In this paper we have explored a class of UV complete theories of massive Galileons, which at low energy are manifestly Galileon invariant, with the exception of the mass term.

Taking candidate low energy theories, we have shown that operators of the form $(\Box \pi)^n$ have the ability to result in Vainshtein screening.
This was suggested by our analytic approximations at small and large $r$.  However, to show that the two asymptotic regimes could indeed be connected to one another, we needed to use numerics.  It turned out that our asymptotic solutions could match and we did not run into any obstacles involving inconsistent boundary conditions or branch cuts.

Generalising the example action \cite{MassGals} to an arbitrary power of self-interaction for the heavy field $H$, we have seen that this class of theories exhibits a massive Galileon symmetry in the light field $\pi$, and that integrating out $H$ only generates terms that respect the symmetry. However, it turns out operators that would normally be neglected in a na\"{\i}ve analysis of the IR equation of motion, due to being suppressed by large powers of the heavy mass $M$, play an important role in determining the behaviour of the solution, and in fact become relevant at a larger radius than operators one might have considered leading order. Interestingly, although individually relevant, these additional operators re-sum to produce a negligible effect, giving a free field profile all the way up to the source radius. Although our candidate low energy theory exhibits screening by virtue of a $(\Box \pi)^n$ operator, in making contact with the UV we necessarily introduce additional operators that entirely disrupt this effect. It is clear that when integrating out a heavy field, a simple truncation is not always sufficient, and in some cases is catastrophically wrong, forcing a careful consideration of all higher order operators being neglected.

With Ref.~\cite{MassGals} having identified a mass term as a potential deformation to avoid positivity bounds in Galileon theories, we investigated its consequences in the context of Wess-Zumino Galileons in the appendix. At first glance the deformation seems to leave the standard screening picture for this theory unaffected, even when one considers loop corrections. However, if we try to connect it to some UV completion and view it from an EFT standpoint, we must necessarily introduce operators that, for heavy enough sources, can dominate over the standard Wess-Zumino terms. Whether this would ruin the screening enjoyed by the deformationless theory or simply increase the radius at which screening occurs is unclear, but the prior results of this paper tell us that the former could be more likely than one might na\"{\i}vely expect. 

\section*{Acknowledgements}

We thank Ben Elder, Peter Millington, and Jonathan Braden for useful discussions during the preparation of this work. CB and DS are  supported by
a Research Leadership Award from the Leverhulme Trust. CB is also supported by a Royal Society University Research Fellowship. AP is supported by a Research Project Grant from the Leverhulme Trust and a STFC consolidated grant. BC and TW are supported by STFC studentships.

\appendix

\section{Massive Wess-Zumino Galileons} \label{WZ}

The familiar Wess-Zumino (WZ) Galileons are a popular modified gravity theory,  first appearing in the context of DGP gravity \cite{DGP, Luty}, giving rise to second-order field equations and Vainshtein screening \cite{galexample}. They are invariant under the standard Galileon symmetry, up to a total derivative and  coupled with the requirement of second-order field equations, this restricts the action to a finite number of operators. These terms can be organised into what are known as the cubic, quartic and quintic Galileons \cite{galexample}.

Despite the many desired features WZ Galileons exhibit, they are impeded from a standard Wilsonian UV completion by the existence of positivity bounds, which restrict the form of low energy scattering amplitudes for scalar theories \cite{posbounds1,posbounds2}. To avoid this limitation, one may deform the theory at low energies, satisfying the bounds, while attempting to keep all other features of the theory intact.

Having shown in section \ref{sec:raise} and \ref{sec:num} that a mass term acts unexpectedly in our candidate theory, we ask whether this type of deformation is acceptable in the context of WZ Galileons, in particular whether the Vainshtein mechanism is preserved. We consider the action,
\begin{equation} \label{WZgals}
S=\int d^4x \left( -\frac{1}{2} \left(\partial \pi \right)^2-\frac{1}{2}m^2 \pi ^2 +\text{WZ terms} + \frac{\pi T}{M_\text{Pl}} \right)
\end{equation}
where the WZ terms are the standard cubic, quartic and quintic Galileons. We will assume for simplicity that the mass term essentially plays no role in screening - its role here is merely to evade the positivity bounds. 

Having posited that simply adding a mass deformation preserves the Vainshtein properties of the theory, while avoiding positivity bounds, we need to consider whether this alteration induces other operators that spoil the screening. Thanks to the Galileon non-renormalisation theorem,  neither the mass or the Wess-Zumino couplings receive radiative corrections \cite{Luty, nonrem1, nonrem2, Galileon_inflation}. However, higher order EFT corrections are of a more general form, which can be written as
\begin{equation}
\frac{(m^2 \pi^2)^a \partial^{2b} (\partial\partial\pi)^c}{\Lambda^{4a+2b+3c-4}}
\end{equation}
where $a,b,c$ are positive integers and we have treated $m^2$ as a spurion. At the level of the equation of motion, this operator yields a term of the form,
\begin{equation}
\mathcal{O} \sim \frac{m^{2a} \partial^{2(b+c)} \pi^{2a+c-1}}{\Lambda^{4a+2b+3c-4}}
\end{equation}
where for the moment we remain agnostic about where the derivatives are operating. Again following the standard procedure, we evaluate the operator on the linearised solution in the static spherically symmetric approximation, for $r\ll m^{-1}$, resulting in
\begin{equation}
\mathcal{O}\bigg\rvert_{\pi_\text{lin}} \sim \frac{m^{2a}}{\Lambda^{4a+2b+3c-4}} m^{2x} \frac{1}{r}^{2(b+c-x)} \left(\frac{\sigma_\text{S}}{r}\right)^{2a+c-1}
\end{equation}
where $x \in [0,b+c]$ and its value depends on the number of $\Box \pi$ insertions in $\mathcal{O}$, and $\sigma_{\text{S}} \equiv M_{\text{S}}/M_{\text{Pl}}$ as in Sec.~\ref{IR}.
Let us now compare this against a standard WZ operator, which looks like,
\begin{equation}
\mathcal{O}_\text{WZ} \sim \frac{(\partial\partial\pi)^L}{\Lambda^{3(L-1)}}
\end{equation}
where $L=2,3,4$. Under the same assumptions, evaluating on the linearised solution gives
\begin{equation}
\mathcal{O}_\text{WZ}\bigg\rvert_{\pi_\text{lin}} \sim \left(\frac{\sigma_{\text{S}}}{r^3 \Lambda^3}\right)^L \Lambda^3\;.
\end{equation}
Comparing the two operators, we obtain a ratio,
\begin{equation} \label{ratio}
\frac{\mathcal{O}}{\mathcal{O}_\text{WZ}}\bigg\rvert_{\pi_\text{lin}} \sim \left(\frac{r_*}{r}\right)^{2a+2b+3c-(1+2x+3L)}
\end{equation}
where the radius $r_*$ at which the two operators become of comparable size, is given by 
\begin{equation}
r_* = \frac{1}{\Lambda} \left(\kappa^{2(a+x)} \sigma_{\text{S}}^{2a+c-L-1}\right)^\frac{1}{2a+2b+3c-(1+2x+3L)}
\end{equation}
where $\kappa \equiv m/\Lambda$ as in Sec.~\ref{IR}.

We know that screening must be contaminated if $\mathcal{O} \gg \mathcal{O}_\text{WZ}$ at $r_\text{V}=\sigma_{\text{S}}^\frac{1}{3} \Lambda^{-1}$, the Vainshtein radius of the WZ theory, as this would mean that when the WZ terms are supposed to start screening, they would be in fact subdominant to the EFT operators.

Setting $\sigma_{\text{S}}\equiv\kappa^{-t}$, we obtain
\begin{equation}
\frac{\mathcal{O}}{\mathcal{O}_\text{WZ}}\bigg\rvert_{\pi_\text{lin}(r_\text{V})} \sim \kappa^P
\end{equation}
where,
\begin{equation}
P = \frac{2}{3}(b+1)t + \frac{2}{3}a(3-2t) + \frac{2}{3}x(3-t)\;.
\end{equation}
If $t \leq \frac{3}{2}$ then $P>0$ and the EFT operators are suppressed relative to the WZ terms. However, if $t>\frac{3}{2}$, then $P$ can be made negative by a sufficiently large choice of $a$. Incidentally, for the parameter values that correspond to the original mass deformation, $P$ is positive all the way up to $t>3$, and so we see that EFT corrections are in general more important.

The value of $t$ is essentially dictated by the size of the source, with heavier sources having a larger $t$. For the Sun, we can estimate $\sigma_\text{S}\sim 10^{39}$, $m\sim H_0$, $\Lambda \sim (1000\text{km})^{-1}$, which gives $t \sim \frac{39}{20}$. We see that, even for a simple example, EFT terms can spoil the screening of the WZ operators.

There is a loophole in the above discussion.  If the Galileon symmetry is only broken by the mass term, then Galileon loops will not generate Galileon breaking operators and we do not obtain arbitrarily high values of $a$. A similar point was already made in \cite{kurt}.  However, the presence of a Galileon breaking interaction, beyond the original mass term, should be enough to generate a full tower of interactions with high values of $a$. Such terms might be expected if the breaking of Galileon symmetry is truly inherited from the UV physics and is present in the couplings between light and heavy fields.

\end{document}